\documentclass[oneside,openany]{book}
\usepackage{graphicx}

\usepackage{etex}
\reserveinserts{28}
\usepackage{natbib}
\usepackage{longtable}
\usepackage{color}
\usepackage{rotating}
\usepackage{floatpag}
\usepackage{pdfpages}
\rotfloatpagestyle{empty}

\usepackage{amsmath}
\makeatletter

\DeclareSymbolFont{AMSa}{U}{msa}{m}{n}

\@ifundefined{checkmark}{%
  \edef\checkmark{\noexpand\mathhexbox{\hexnumber@\symAMSa}58}
}{}
\@ifundefined{circledR}{%
  \edef\circledR{\noexpand\mathhexbox{\hexnumber@\symAMSa}72}
}{}

\makeatother

\usepackage{amsthm}
\usepackage{graphicx}

\usepackage{makeidx}

\usepackage{morefloats}

\DeclareFontEncoding{TS1}{}{}
\DeclareTextCommandDefault{\textcopyright}{{\fontencoding{TS1}\fontfamily{\rmdefault}\selectfont\char"A9}}

\graphicspath{{../1_Art/}}

\AtBeginDocument{%

}

\makeatletter

\DeclareRobustCommand{\nu}{\text{\usefont{OML}{npxmi}{m}{it}\selectfont\char23}}

\DeclareSymbolFont{normalalphabets}{OML}{cmm}{m}{it}%
\SetSymbolFont{normalalphabets}{bold}{OML}{cmm}{b}{it}%
\DeclareMathSymbol{v}{\mathalpha}{normalalphabets}{`v}%

\makeatother

\begin{document}

\frontmatter

\pagenumbering{roman}
\setcounter{page}{1}

\begin{figure}[p]
    \vspace*{-4cm}
    \hspace*{-4.1cm}
        \includegraphics[angle=0,width=1.7\textwidth]{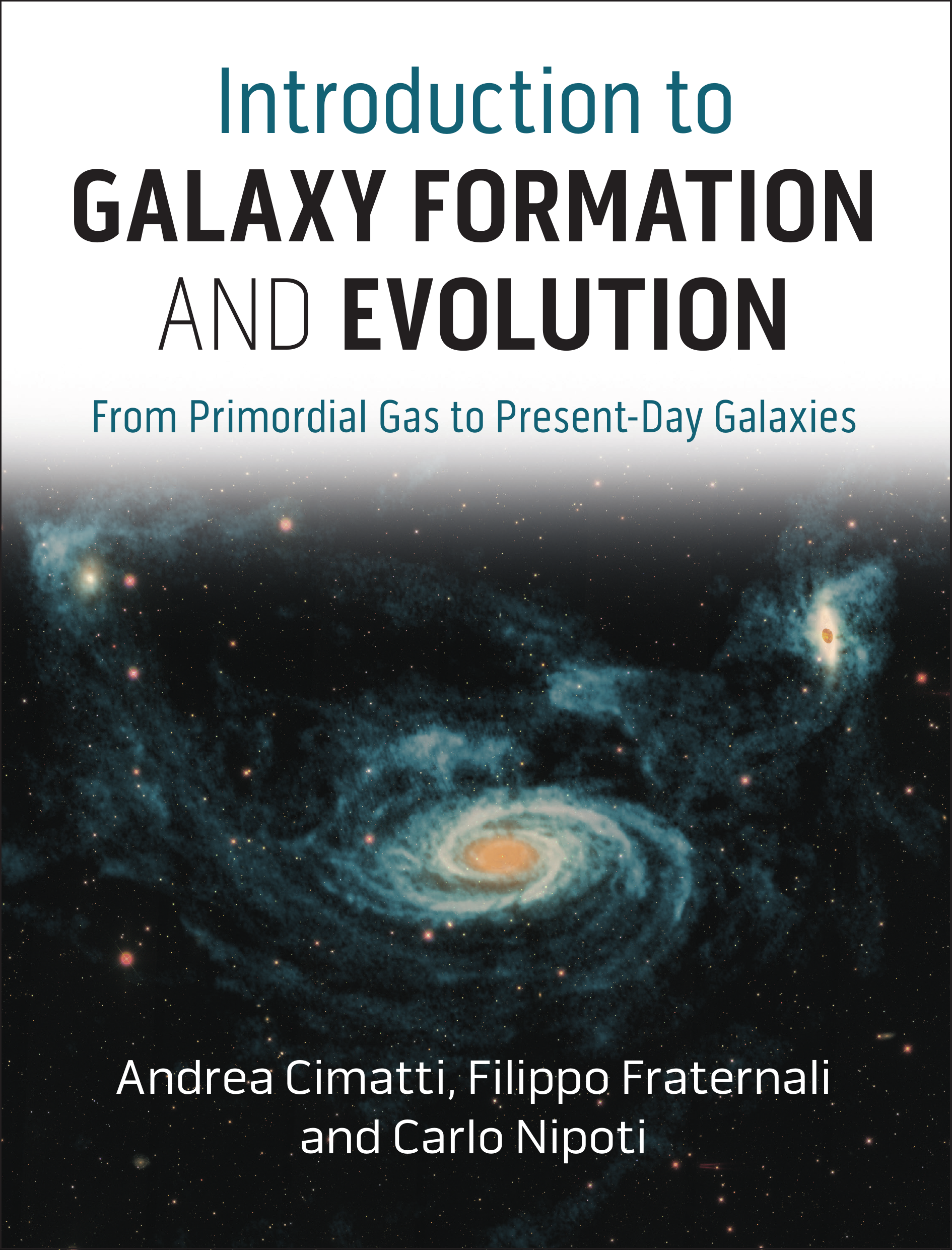}
\end{figure}

\begin{figure}[p]
    \vspace*{-4cm}
    \hspace*{-4.1cm}
        \includegraphics[width=1.7\textwidth]{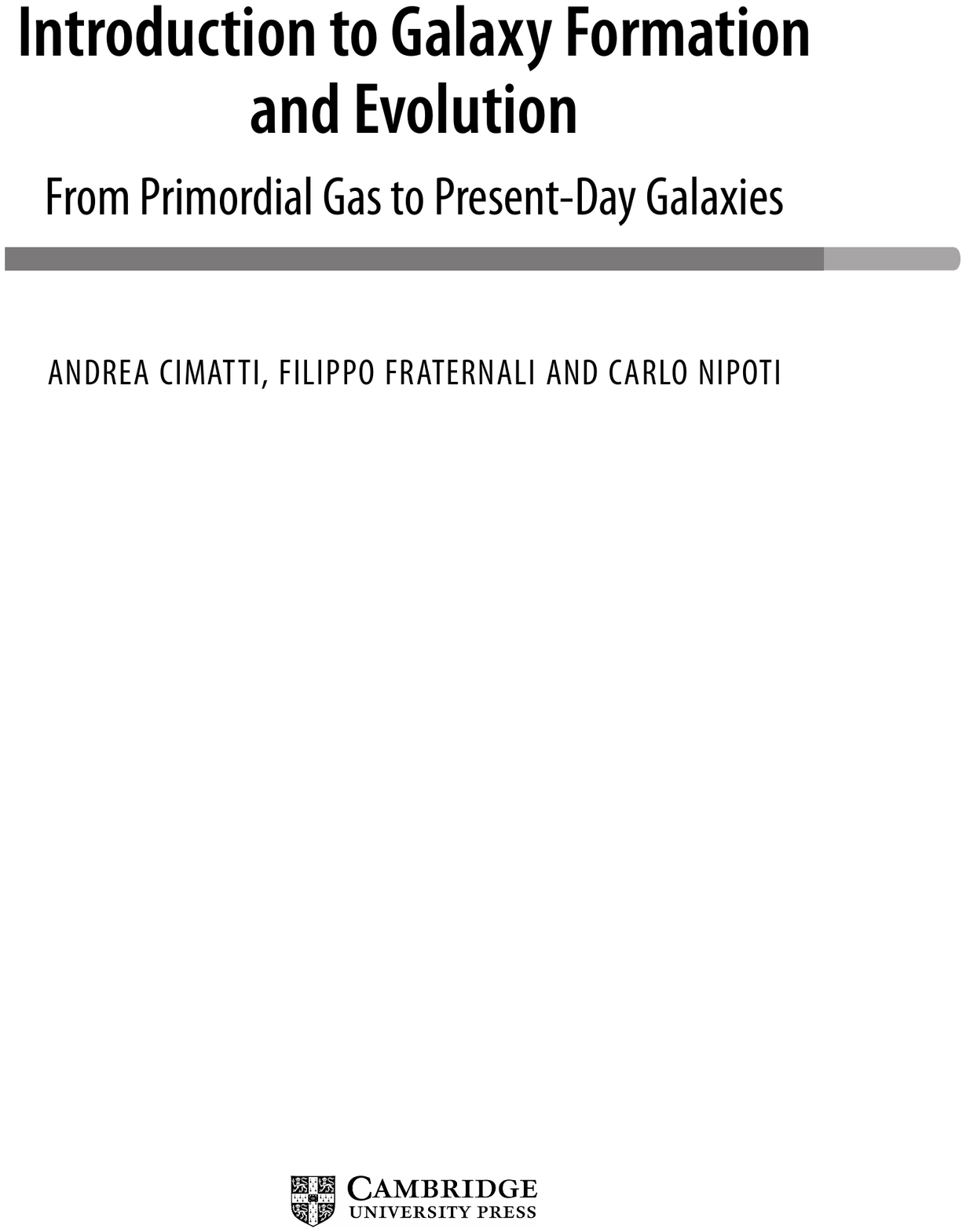}
\end{figure}


\addcontentsline{boc}{bschapter}{Preface}
\addcontentsline{toc}{fmschapter}{Preface}

\makeatletter
\@openrightfalse
\makeatother

\chapter*{Preface}
\label{sec:preface}

{\bf Why This Book?}
\\
The study of galaxy formation and evolution is one of the most active and fertile
fields of modern astrophysics. It also covers a wide range of topics intimately
connected with cosmology and with the evolution of the Universe as a whole. The key to
decipher galaxy formation and evolution is to understand the complex physical processes
driving the evolution of ordinary matter during its gravitational interplay with
dark matter halos across cosmic time. The central theme is therefore how galaxies
formed and developed their current properties starting from a diffuse distribution of
gas in the primordial Universe.
This research field
requires major efforts in the observation of galaxies over a wide range of distances,
and in the theoretical modelling of their formation and evolution. The synergy between
observations and theory is therefore essential to shed light on how galaxies formed and
evolved. In the last decades, both observational and theoretical
studies have undergone rapid developments. The availability of new
telescopes operating from the ground and from space across the entire electromagnetic
spectrum opened a new window on distant galaxies. At the same time, major observational
campaigns, such as the Sloan Digital Sky Survey, provided huge samples of
galaxies in the present-day Universe with unprecedented statistics and allowed
one to define the `zero-point' for evolutionary studies. In parallel, the theoretical
models experienced a major advance thanks to the improved
performance of numerical simulations of galaxy formation within the cosmological
framework.

The idea for this book originated from the difficulties we faced when teaching our
courses. We lacked a single and complete Master-level student textbook on how
galaxies formed and evolved. This textbook aims to fill a gap between highly
specialised and very introductory books on these topics, and enables students to easily find the required
information in a single place, without having to consult many sources.


The aim of the book is twofold. The first is to provide an introductory, but complete,
description of the key physical processes that are important in galaxy
formation and evolution, from the primordial to the present-day Universe. The second is
to illustrate what physical and evolutionary information can be derived using
multi-wavelength observations. As the research field of galaxy formation and evolution
is relatively young and rapidly evolving, we do not attempt to give a complete review
of all topics, but rather we try to focus on only the most solid results.

\vskip\baselineskip

\noindent{\bf Readership and Organisation}

\noindent This textbook assumes a background in general physics at the Bachelor level, as
well as in introductory astronomy, fundamentals of radiative processes in
astrophysics, stellar evolution and the fundamentals of hydrodynamics.
Although this book is primarily intended for students at Master degree level, it can
be used as a complement to Bachelor-level courses in extragalactic astrophysics,
and we think it can also be a valuable guide to PhD students and researchers.

The content of the chapters is organised as follows. After a general introduction
to the field of galaxy formation and evolution (Chapter 1), the book starts with a
brief overview on the cosmological framework in which galaxies are placed (Chapter 2).
The aim of this chapter is to provide the reader with the key information useful for
the rest of the textbook: the Big Bang model, the expansion of the Universe, redshift,
the look-back time, the cosmological parameters and the matter--energy cosmic budget.
Then, the book continues with a set of four chapters dedicated to the properties of
present-day galaxies seen as the endpoint of the evolution that has occurred during the time
frame spanned by the age of the Universe ($\approx$13.8 billion years). In particular,
Chapter 3 illustrates the statistical properties of galaxies (e.g. morphologies, sizes,
luminosities, masses, colours, spectra) and includes a description of active galactic nuclei. The structure,
components and physical processes of star-forming and early-type galaxies are presented
in Chapter 4 and Chapter 5, respectively. Chapter 4 includes also a description of our
own Galaxy seen as a reference benchmark when studying the physics of star-forming
galaxies from the `inside' and with a level of detail not reachable in external galaxies.
Chapter 6 deals with the influence of the environment on galaxy properties, and with
the spatial distribution of galaxies on large scales. Then, Chapter 7 focuses on the
general properties of dark matter halos, and their hierarchical assembly across cosmic
time: these halos are crucial because they constitute the skeleton where
galaxy formation takes place. Chapter 8 deals with the main `ingredients' of galaxy
formation theory through the description of the key physical processes determining the
evolution of baryons within dark matter halos (e.g. gas cooling and heating, star
formation, chemical evolution, feedback processes).
The subsequent Chapter 9 is dedicated to the evolution of primordial baryonic matter
in the early Universe, from the cosmological recombination to the formation of the
first luminous objects a few hundred million years after the Big Bang, and the
consequent epoch of reionisation. Chapter 10 provides a general description of the
theoretical models of the formation and evolution of different types of galaxies,
including an introduction to the main methods of numerical modelling of galaxy formation.
Finally, Chapter 11 presents a general overview of galaxy evolution based on the
direct observation of distant galaxies and their comparison with present-day galaxy
types.


\vskip\baselineskip

\noindent {\bf References}

\noindent As of writing this book, there are tens of thousands of refereed papers in the literature
on galaxy formation and galaxy evolution; not to mention several books on galaxies
and cosmology available on the market. This implies that choosing the most significant
references for a book like this is really challenging. The difficulty is exacerbated
by the very fast evolution of this research field. For these reasons, our choice
has been pragmatic and minimalistic. We excluded references before 1900, and we decided
to reduce as much as possible the citations to research articles (including our own papers),
unless they present a major discovery or a turning point for a given topic, or they are
particularly useful for students. Instead, we much preferred to cite recent review
articles because they provide an introductory and as much as possible
unbiased source of information that is more suitable for students. However, also in
this case, it was not feasible to cite all the reviews available in the literature.
In the same spirit, the figures selected from the literature were chosen based on
their clarity and usefulness to students.
Finally, we also suggested a few books where readers can find more
details on several topics treated in this textbook. The obvious consequence is that the
reference list is unavoidably incomplete. We apologise to any author whose
publications may have been overlooked with the selection approach that we adopted.

\vskip\baselineskip

\noindent {\bf Acknowledgements}

%
%
\noindent This book has benefited from the input of colleagues and students
who have helped us in a variety of different and crucial ways. Many of the
figures in this book have been produced {\em ad hoc} for us. We are grateful
to the authors of these figures, to whom we give credit in
the captions. Here we also wish to explicitly thank our colleagues who have taken
the time to read parts of the text, and/or gave us comments and advice that were fundamental to improve the quality of the book.
These are: Lucia Armillotta, Ivan Baldry, Matthias Bartelmann, James Binney, Fabrizio
Bonoli, Fabio Bresolin, Volker Bromm, Marcella Brusa, Luca Ciotti, Peter Coles, Romeel
Dav\'{e}, Gabriella De Lucia, Mark Dickinson, Enrico Di Teodoro, Elena D'Onghia, Stefano
Ettori, Benoit Famaey, Annette Ferguson, Daniele Galli, Roberto Gilli,
Amina Helmi, Giuliano Iorio, Peter Johansson, Inga Kamp, Amanda Karakas, Rob Kennicutt, Dusan Kere\v{s}, Leon Koopmans,
Mark Krumholz, Federico Lelli,
Andrea Macci\`{o}, Mordecai Mac Low, Pavel Mancera Pi\~na, Antonino Marasco, Claudia Maraston, Federico Marinacci, Davide
Massari, Juan Carlos Mu\~noz-Mateos, Kyle Oman, Tom Oosterloo, Max Pettini,
Gabriele Pezzulli, Anastasia Ponomareva, Lorenzo Posti, Mary Putman, Sofia Randich,
Alvio Renzini, Donatella Romano, Alessandro Romeo, Renzo Sancisi, Joop Schaye,
Ralph Sch\"{o}nrich, Mattia Sormani, Eline Tolstoy, Scott Tremaine, Tommaso Treu,
Mark Voit, Marta Volonteri, Jabran Zahid, Gianni Zamorani and Manuela Zoccali.

\index{21 cm line|seealso{neutral atomic gas}}
\index{AGB|see{asymptotic giant branch}}
\index{AGNs|see{active galactic nuclei}}
\index{AMD|see{angular momentum distribution}}
\index{AMR|see{adaptive mesh refinement}}
\index{Andromeda|see{M\,31}}
\index{angular momentum!specific|see{specific angular momentum}}
\index{atomic gas|see{neutral atomic gas {\em or} ionised gas}}
\index{bands|see{filters}}
\index{BAOs|see{baryon acoustic oscillations}}
\index{BCGs|see{brightest cluster galaxies}}
\index{BCDs|see{blue compact dwarf galaxies}}
\index{BHAR|see{black hole accretion rate}}
\index{boxy bulges|see{bulges, peanut-shaped}}
\index{BLRs|see{broad-line regions}}
\index{black hole mass--bulge mass relation|see{Magorrian relation}}
\index{black holes!supermassive|see{supermassive black holes}}
\index{BPT|see{Baldwin--Phillips--Terlevich}}
\index{bubbles!in X-rays|see{cavities in X-rays}}
\index{clouds!molecular|see{molecular clouds}}
\index{clusters of stars|see{globular clusters {\em or} open clusters}}
\index{CO|see{carbon monoxide}}
\index{cooling!radiative|see{radiative cooling}}
\index{CMB|see{cosmic microwave background}}
\index{core-S\'ersic galaxies|see{cored galaxies}}
\index{core-collapse supernovae|see{supernovae, Type II}}
\index{CDM|see{cold dark matter}}
\index{CGM|see{circumgalatic medium}}
\index{CMDs|see{colour--magnitude diagrams}}
\index{CNM|see{cold neutral medium}}
\index{CSPs|see{composite stellar populations}}
\index{dark matter!cold|see{cold dark matter}}
\index{dark matter!hot|see{hot dark matter}}
\index{dark matter!warm|see{warm dark matter}}
\index{dEs|see{dwarf galaxies, elliptical}}
\index{disc galaxies|see{star-forming galaxies}}
\index{discs|seealso{star-forming galaxies}}
\index{diffuse matter|see{interstellar medium}}
\index{dIrrs|see{dwarf galaxies, irregular}}
\index{distant galaxies!selection|see{selection of distant galaxies}}
\index{DLAs|see{damped Lyman-$\alpha$ absorbers}}
\index{dropout galaxies|see{Lyman-break galaxies}}
\index{dSphs|see{dwarf galaxies, spheroidal}}
\index{effective models|see{subgrid prescriptions}}
\index{elliptical galaxies|seealso{early-type galaxies}}
\index{encounters of galaxies!high-speed|see{fly-bys}}
\index{EoR|see{epoch of reionisation}}
\index{EROs|see{extremely red objects}}
\index{evolution!chemical|see{chemical evolution}}
\index{evolution!secular|see{secular evolution}}
\index{feedback!stellar|see{stellar feedback}}
\index{stellar feedback!fountains|see{galactic fountains}}
\index{free-free emission|see{bremsstrahlung}}
\index{FR|see{Fanaroff--Riley}}
\index{fountains|see{galactic fountains}}
\index{galaxies!dwarf|see{dwarf galaxies}}
\index{galaxies!disc|see{star-forming galaxies}}
\index{galaxies!late-type|see{star-forming galaxies}}
\index{galaxies!star-forming|see{star-forming galaxies}}
\index{galaxies!spiral|see{star-forming galaxies}}
\index{gas!atomic|see{neutral atomic gas {\em or} ionised gas}}
\index{gas!collisionally ionised|see{collisionally ionised gas}}
\index{gas!in clusters of galaxies|see{intracluster medium}}
\index{gas!in galaxies|see{interstellar medium}}
\index{gas!ionised|see{ionised gas}}
\index{gas!neutral atomic|see{neutral atomic gas}}
\index{gas!photoionised|see{photoionised gas}}
\index{gas!primordial|see{primordial gas}}
\index{gas!molecular|see{molecular gas}}
\index{gas streams|seealso{filaments!of gas}}
\index{gas filaments|see{filaments!of gas}}
\index{GMCs|see{giant molecular clouds}}
\index{halo gas|see{circumgalactic medium}}
\index{halo quenching|see{mass quenching}}
\index{halos!dark|see{dark matter halos}}
\index{H\,I|see{neutral atomic gas}}
\index{H$_2$|see{molecular hydrogen}}
\index{high-speed encounters|see{fly-bys}}
\index{HIM|see{hot intracloud medium}}
\index{hydrogen!neutral atomic|see{neutral atomic gas}}
\index{hydrogen!molecular|see{molecular hydrogen}}
\index{hydrogen!ionised|see{ionised gas}}
\index{IMF|see{initial mass function}}
\index{internal quenching|see{mass quenching}}
\index{interstellar dust|see{dust}}
\index{interstellar medium!molecular|see{molecular gas {\em or} molecular hydrogen}}
\index{IFUs|see{integral field units}}
\index{IRA|see{instantaneous recycling approximation}}
\index{Kennicutt law|see{Schmidt--Kennicutt law}}
\index{K+A systems|see{E+A systems}}
\index{late-type galaxies|see{star-forming galaxies}}
\index{LBGs|see{Lyman-break galaxies}}
\index{LF|see{luminosity function}}
\index{LOSVD|see{line-of-sight velocity distribution}}
\index{LTGs|see{star-forming galaxies}}
\index{Galactic|see{Milky Way}}
\index{galaxies!at high redshift|see{distant galaxies}}
\index{galaxies!distant|see{distant galaxies}}
\index{galaxies!first|see{first galaxies}}
\index{galaxies!starburst|see{starburst galaxies}}
\index{Galaxy|see{Milky Way}}
\index{GRBs|see{gamma-ray bursts}}
\index{half-light radius|see{effective radius}}
\index{Hickson compact groups|see{compact groups}}
\index{HDM|see{hot dark matter}}
\index{HSBs|see{high surface brightness galaxies}}
\index{Hubble sequence|see{Hubble classification}}
\index{ICM|see{intracluster medium}}
\index{IGM|see{intergalactic medium}}
\index{ISM|see{interstellar medium}}
\index{lensing|see{gravitational lensing}}
\index{LLSs|see{Lyman-limit systems}}
\index{LMC|see{Large Magellanic Cloud}}
\index{LSBs|see{low surface brightness galaxies}}
\index{LSS|see{large-scale structure}}
\index{merging|see{merger {\em or} mergers}}
\index{mergers!dissipationless|see{mergers, dry}}
\index{mergers!dissipative|see{mergers, wet}}
\index{metal abundance|see{metallicity}}
\index{migration|see{radial migration}}
\index{molecular clouds!giant|see{giant molecular clouds}}
\index{MOND|see{modified Newtonian dynamics}}
\index{NFW|see{Navarro--Frenk--White}}
\index{nightglow|see{airglow}}
\index{NLRs|see{narrow-line regions}}
\index{fundamental observers|see{comoving observers}}
\index{passive galaxies|see{early-type galaxies}}
\index{PAHs|see{polycyclic aromatic hydrocarbons}}
\index{photo-$z$|see{photometric redshift}}
\index{PISNe|see{pair instability supernovae}}
\index{phase-space density|see{distribution function, phase-space}}
\index{photoheating|see{photoionisation heating}}
\index{power-law galaxies|see{coreless galaxies}}
\index{power spectrum!cosmological|see{cosmological power spectrum}}
\index{stars!Population I|see{Population I stars}}
\index{stars!Population II|see{Population II stars}}
\index{stars!Population III|see{Population III stars}}
\index{preheating|see{feedback, preventive}}
\index{PSF|see{point spread function}}
\index{QSOs|see{quasars}}
\index{quasi-stellar objects|see{quasars}}
\index{quiescent galaxies|see{early-type galaxies}}
\index{recipes|see{prescriptions}}
\index{reddening|see{colour excess {\em and} dust}}
\index{ripples|see{shells}}
\index{Roche radius|see{Hill radius}}
\index{satellite quenching|see{environmental quenching}}
\index{S0 galaxies|see{lenticular galaxies}}
\index{lenticular galaxies|seealso{early-type galaxies}}
\index{SAMs|see{semi-analytic models}}
\index{scaling laws|see{scaling relations}}
\index{simulations!cosmological|see{cosmological simulations}}
\index{stars!first|see{first stars}}
\index{stripping|see{ram-pressure stripping {\em or} tidal stripping}}
\index{S\'ersic galaxies|see{coreless galaxies}}
\index{S\'ersic profile|see{S\'ersic law}}
\index{SED|see{spectral energy distribution}}
\index{SFGs|see{star-forming galaxies}}
\index{SFMS|see{star formation main sequence}}
\index{SFGs|see{star-forming galaxies}}
\index{SFR|see{star formation rate}}
\index{SHMR|see{stellar-to-halo mass relation}}
\index{SLED|see{spectral line energy distribution}}
\index{SMBHs|see{supermassive black holes}}
\index{SMC|see{Small Magellanic Cloud}}
\index{SMGs|see{submillimetre galaxies}}
\index{SMF|see{stellar mass function}}
\index{SNe|see{supernovae}}
\index{SNRs|see{supernova remnants}}
\index{SPH|see{smoothed particle hydrodynamics}}
\index{spiral galaxies|see{star-forming galaxies}}
\index{SPS models|see{stellar population synthesis models}}
\index{sSFR|see{specific star formation rate}}
\index{SSPs|see{simple stellar populations}}
\index{spheroids|seealso{early-type galaxies}}
\index{star-forming galaxies|seealso{discs}}
\index{star-forming galaxies!fountains|see{galactic fountains}}
\index{stars!feedback|see{stellar feedback}}
\index{streams!of gas|seealso{filaments of gas}}
\index{Sun|see{solar}}
\index{subresolution recipes|see{subgrid prescriptions}}
\index{supernova feedback|see{stellar feedback}}
\index{SZE|see{Sunyaev--Zeldovich effect}}
\index{TFR|see{Tully--Fisher relation}}
\index{UFDs|see{ultra-faint dwarf galaxies}}
\index{ULIRGs|see{ultra-luminous infrared galaxies}}
\index{UVB|see{ultraviolet background}}
\index{WDM|see{warm dark matter}}
\index{WHIM|see{warm--hot intergalactic medium}}
\index{WIM|see{warm ionised medium}}
\index{WIMPs|see{weakly interacting massive particles}}
\index{WNM|see{warm neutral medium}}

\cleardoublepage

\mainmatter


\pagenumbering{arabic}
\setcounter{page}{1}

%
\addtocontents{boc}{\protect\vskip0.5\baselineskip}

\makeatletter
\@openrightfalse
\makeatother

\chapter{Introduction}
\label{sec:introduction}

\section{Galaxies: a Very Brief History}
\label{sec:chap1_galaxies_history}

Galaxies are gravitationally bound systems made of stars,
interstellar matter (gas and dust), stellar remnants (white dwarfs, neutron stars
and black holes) and a large amount of dark matter. They are varied systems with
a wide range of morphologies and properties. For instance, the characteristic sizes
of their luminous components are from $\sim$ 0.1 kpc to tens of kiloparsecs, whereas the
optical luminosities and stellar masses are in the range $10^3$--$10^{12}$
in solar units. Roughly spherical halos of dark matter dominate the mass budget of galaxies.
As a reference, the size of the stellar disc of our Galaxy\footnote{The terms
Galaxy (with the capital G) or Milky Way are used to indicate the galaxy where the
Sun, the authors and the readers of this book are located.} is about 20 kpc,
whereas the dark matter halo is thought to be extended out to $\approx$300 kpc. The
total mass of the Galaxy, including dark matter, is $\sim 10^{12}$ $\mathcal{M}_{\odot}$,
whereas the stellar and gas masses amount to only $\approx 5\times 10^{10}$ $\mathcal{M}_{\odot}$
and $\approx 6 \times 10^{9}$ $\mathcal{M}_{\odot}$, respectively.

The discovery of galaxies (without knowing their nature) dates back to when the
first telescope observations showed the presence of objects, originally called
nebulae, whose light appeared diffuse and fuzzy. The first pioneering observations
of these nebulae were done with telescopes by C.\ Huygens in the mid-seventeenth century, and by
E.\ Halley and N.-L.\ de Lacaille in the first half of the eighteenth century. Interestingly,
in 1750, T.\ Wright published a book in which he interpreted the Milky Way as a
flat layer of stars and suggested that nebulae could be similar systems at
large distances. The philosopher I.\ Kant was likely inspired by
these ideas to the extent that, in 1755, he explained that these objects (e.g.\ the Andromeda
galaxy) appear nebulous because of their large distances which make it impossible
to discern their individual stars. In this context, the Milky Way was interpreted as
one of these many stellar systems (island universes).

In 1771, C.\ Messier started to catalogue the objects which appeared
fuzzy based on his telescope observations. These objects were identified by the
letter M (for Messier) followed by a number. Now we know that some of these
objects are located within our Galaxy (star clusters and emission nebulae;
e.g.\ M\,42 is the Orion nebula), but some are nearby galaxies bright enough to be visible
with small telescopes (e.g.\ M\,31 is the Andromeda galaxy). However, Messier did not
express any opinion about the nature and the distance of these systems.
Since late 1700, W.\ Herschel, C.\ Herschel and J.\ Herschel increased the sample
of diffuse objects thanks to their larger telescopes, and classified them depending on
their observed features. In 1850, W.\ Parsons (Lord Rosse) noticed that
some of these nebulae exhibited a clear spiral structure (e.g.\ M\,51).

Since late 1800, the advent of astronomical photography allowed more detailed
observations to be performed, and these studies triggered a lively discussion about the nature of the
spiral nebulae and their distance. This led to the so-called Great Debate, or the
\hbox{Shapley--Curtis} debate referring to the names of the two astronomers who, in 1920,
proposed two widely different explanations about spiral nebulae. On the one hand,
H. Shapley argued that these objects were interstellar gas clouds located within one
large stellar system. On the other hand, according to H.\ Curtis, spiral nebulae were external
systems, and our Galaxy was one of them.
Clearly, this debate involved not only the very nature of these objects, but also
the size and the extent of the Universe itself. The issue was resolved soon
after with deeper observations.
In 1925, using the 100-inch telescope at Mount Wilson Observatory,
E.\ Hubble identified individual stars in M 31 and M 33 and discovered variable stars
such as Cepheids and novae. In particular, Cepheids are pulsating giant stars
that can be exploited as distance indicators. These stars are what astronomers
call `standard candles', i.e.\ objects whose intrinsic luminosity is known {\em a priori},
and that therefore can be used to estimate their distance. In 1912,
it was found by H.\ Leavitt that the intrinsic luminosity of Cepheids is proportional to
the observed period of their flux variation. Thus, once the period is measured, the intrinsic
luminosity is derived and, therefore, the distance can be estimated. Based on these
results, Hubble demonstrated that spiral nebulae were at very large distances, well
beyond the size of our Galaxy, and that therefore they were indeed external galaxies.


The term `galaxy' originates from the Greek $\gamma \acute{\alpha} \lambda \alpha$, which means
milk, and it refers to the fuzzy and `milky' appearance of our own Milky Way when observed with
the naked eye. Also external galaxies look `milky' when observed with small telescopes.
Discovering that galaxies were external systems also implied that the Universe was
much larger than our Galaxy, and this was crucial to open a new window on cosmology
in general. In modern astrophysics, the term `nebula' is still used, but it refers only to
objects within the interstellar medium of galaxies. Notable examples are the
emission nebulae where the gas is photoionised by hot massive stars, dark nebulae
which host cold and dense molecular gas mixed with interstellar dust, and planetary
nebulae produced by the gas expelled by stars with low to intermediate mass during
their late evolutionary phases. Since the discovery of Hubble, spiral nebulae have
therefore been called spiral galaxies. In 1927--1929, based on galaxy samples for which
radial velocities and distances were available, G.\ Lema\^itre and Hubble found that
galaxies are systematically receding from us. In particular, their radial velocity is
proportional to their distance: the farther away the galaxies, the higher the redshift
of their spectral lines, and therefore the velocity at which they move away from us.
This crucial discovery led to the Hubble--Lema\^itre law\footnote{In October
2018, the members of the International Astronomical Union (IAU) voted and recommended
to rename the Hubble law as the Hubble--Lema\^itre law.} which is the
experimental proof that the Universe is expanding.

\section{Galaxies as Astrophysical Laboratories}
\label{sec:chap1_galaxies_laboratories}

Present-day galaxies display a variety of properties and span a very
broad range of luminosities, sizes and masses. At first sight, this already
suggests that galaxy formation and evolution is not a simple
process. However, the existence of tight scaling relations involving galaxy
masses, sizes and characteristic velocities (e.g.\ the Tully--Fisher relation
and the fundamental plane) indicates
some regularities in the formation and assembly of these systems.


\begin{figure}[!b]
{\includegraphics{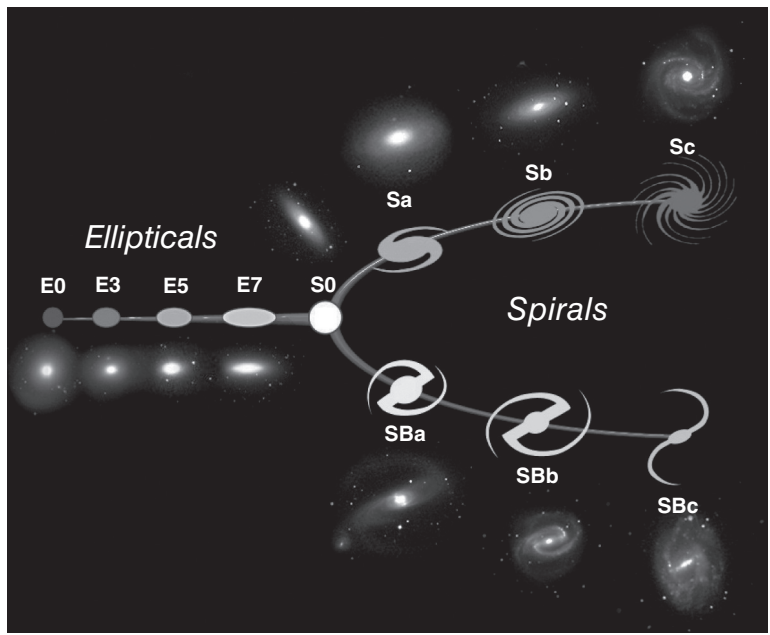}}
{\caption{The Hubble classification of galaxy morphology. \textcopyright\ NASA and ESA,
reproduced with permission.}\label{fig:intro_hubblefork}}
\end{figure}

The first distinctive feature of a galaxy is its morphology. The shape of a
galaxy as observed on the sky plane is a combination of the intrinsic
three-dimensional (3D) structure and its orientation relative to the line of
sight. Present-day galaxies show a broad range of shapes.
Understanding the physical formation and evolution of the morphological types
remains one of the most important, and still open, questions in extragalactic astrophysics.
The first systematic study in the optical waveband dates back to 1926, when Hubble
started a classification of galaxy morphologies following an
approximate progression from simple to complex forms. In particular,
Hubble proposed a tuning fork diagram
on which the main galaxy types can be placed. Based on this classification,
galaxies were divided into three main classes: ellipticals, lenticulars and
spirals, plus a small fraction of irregulars. As shown in
{Fig.}\ \ref{fig:intro_hubblefork}, the Hubble sequence starts from the left
with the class of ellipticals~(E). This class
is further divided into subclasses as a function of their observed
flattening. Perfectly round ellipticals are called E0, whereas the
most flattened are the E7. If the observed shape of these
galaxies is approximated by ellipses, their flattening is
related to the ellipticity $\epsilon = (a-b)/a$, where $a$ and $b$ are
the observed semi-major and semi-minor axes, respectively. The number written after
the letter E is the integer closest to $10\epsilon$.  Proceeding
beyond the E7 class, galaxies start to display morphologies with a
central dominant spheroidal structure (the so-called bulge) surrounded by a
fainter disc without spiral arms. These systems are classified as lenticulars
(S0) and represent a morphological transition from ellipticals
to spirals. Proceeding further to the right, the tuning fork is
bifurcated in two prongs populated by the two main classes of
spiral (S) galaxies. In both prongs, spirals have
the common characteristic of having a disc-like appearance with well defined
spiral arms originating from the centre and extending throughout the outer
regions. The top prong includes the so-called normal
spirals characterised by a central bulge surrounded by a disc.
These spirals are classified Sa, Sb and Sc as a function of decreasing
prominence of the bulge (with respect to the disc) and increasing importance of
the spiral arms. The bottom prong includes the barred spirals
(SB) which show a central bar-like structure which connects the bulge
with the regions where the spiral arms begin. Moving further to the
right, i.e.\ beyond Sc types, all galaxies not falling into the previous classes
are classified as irregulars (Irr).

Subsequent studies showed that
ellipticals and lenticulars are red systems, made of old stars, with weak or
absent star formation, high stellar masses, with a wide range of kinematic properties
(from fast to absent rotation), and preferentially located in
regions of the Universe where the density of galaxies is higher. On the other
side of the tuning fork, spirals are bluer, have ongoing star formation,
larger fractions of cold gas, stellar populations with a wide range of ages,
kinematics dominated by rotation, and are found preferentially in regions with
lower density of galaxies.
Given the wide range of properties displayed by present-day galaxies, it is
crucial to investigate the physical processes which led to their formation
and evolution. The study of galaxies involves a wide range of galactic
and sub-galactic scales ranging from hundreds of kiloparsecs down to sub-parsec level
depending on the processes that are considered. In this respect, galaxies can be seen as
`laboratories' where a plethora of astrophysical processes can be investigated.

\section{Galaxies in the Cosmological Context}
\label{sec:chap1_galaxies_cosmocontext}

Besides their role as astrophysical laboratories, galaxies can be placed in a broader
context and exploited as point-like
luminous `particles' which trace the distribution of matter on scales much larger
than the size of individual galaxies. This distribution, called large-scale
structure, is the 3D spatial distribution of matter in the Universe
on scales from tens of megaparsecs to gigaparsecs. Due to its characteristic shape, the large-scale
structure is also called the cosmic web.
The study of galaxies on these large scales has deep connections with cosmology, the
branch of physics and astrophysics that studies the general properties, the
matter--energy content and the evolution of the Universe as a whole.
Modern cosmology rests on two major observational pillars. The first is
the expansion of the Universe (Hubble--Lema\^itre law). The second is the nearly uniform
radiation background observed in the microwaves, the cosmic microwave background
(CMB), discovered in 1965 by A. Penzias and R. Wilson. The spectrum of the CMB is an
almost perfect black body with a temperature $T\ \simeq\ 2.726$ K.
The CMB radiation is interpreted as
the thermal relic of the Big Bang that occurred about 13.8 Gyr ago when the Universe
originated as a hot plasma with virtually infinite temperature and density.
Although the detailed properties of the Big Bang itself are unknown, an expanding Universe can
be described using the Einstein equations of general relativity together
with the Friedmann--Lema\^itre--Robertson--Walker metric. The current view of the Universe
relies on the Big Bang model and on the so-called $\Lambda$CDM cosmological framework.
In this scenario, also known as standard cosmology, the Universe is homogeneous and
isotropic on large scales, and it is made of ordinary matter (i.e.\ baryonic matter),
neutrinos, photons and a mysterious component of cold dark matter (CDM).
CDM is dominant with respect to ordinary matter as it amounts to about 84\% of the whole
matter present in the Universe. CDM is thought to be composed of non-relativistic massive
particles that interact with each other and with ordinary matter only through the
gravitational force. However, the nature and individual mass of these particles are
currently unknown. For this reason, this is one of the main open questions of modern physics.
In addition, a further component, called dark energy, is required to explain
the current acceleration of the Universe expansion that S. Perlmutter, B. Schmidt and
A. Riess discovered in 1998 exploiting distant supernovae as standard candles.
In standard cosmology, the space-time geometry is flat
(Euclidean), and dark energy is assumed to be a form of energy density (known as vacuum
energy) which is constant in space and time. This form of dark energy is indicated by
$\Lambda$ and called the cosmological constant. However, other possibilities
(e.g.\ a scalar field) are not excluded, and the nature of dark energy is currently unknown.
This represents another big mystery of modern physics.


The $\Lambda$CDM model can be fully described by a small number of quantities called
cosmological parameters which measure the relative fractions of the matter--energy
components and constrain the geometry of the Universe. The $\Lambda$CDM model is now
supported by a variety of cosmological probes such as the CMB, the Hubble expansion rate
estimated from Type Ia supernovae, the properties of the large-scale structure and
the mass of galaxy clusters. If the $\Lambda$CDM model is assumed, the observational
results constrain the cosmological parameters with extremely high accuracy. In particular,
in the present-day Universe, dark energy contributes $\approx$70\% of the matter--energy
budget of the Universe, whereas the contributions of dark matter and baryons
amount to $\approx$25\% and $\approx$5\%, respectively, plus a negligible fraction of photons
and neutrinos. The relative uncertainties on these fractions are very small (sub-per cent
level). For this reason, modern cosmology is also called precision cosmology.
However, it remains paradoxical that the nature of dark matter and dark energy,
which together make 95\% of the Universe, is still completely unknown despite the accuracy
with which we know their relative importance.

Once the cosmological framework has been established, present-day galaxies can be
seen as the endpoints that enclose crucial information on how baryonic and dark matter
evolved as a function of cosmic time. In this regard, galaxies are also useful to test
the $\Lambda$CDM cosmology. For instance, the current age of the oldest stars in
galaxies should not be older than the age of the Universe itself, estimated to be
about 13.8 Gyr based on observational cosmology. This key requirement is met by
the age estimates of the Galactic globular clusters based on the Hertzsprung--Russell
diagram.

\begin{figure}[!b]
{\includegraphics{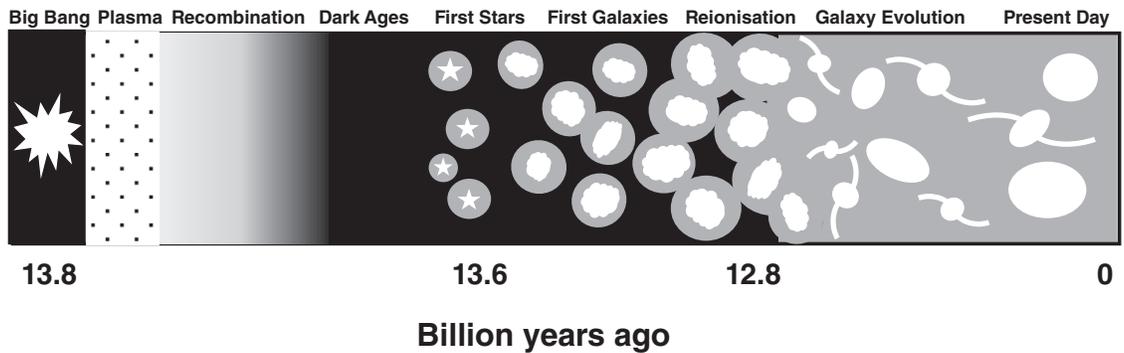}}
{\caption{A sketch of the main epochs which characterised the evolution of the
Universe, starting from the Big Bang. After the formation of the first stars
and galaxies, galaxies followed different evolutionary paths which led to the
assembly of the galaxy types that we observe in the present-day Universe.}
\label{fig:ch1_cosmic_timeline}}
\end{figure}

\section{Galaxies: from First Light to Present-Day Galaxies}
\label{sec:chap1_galaxies_evolution}

Galaxies were originated from the primordial gas present in the early Universe.
{Fig.}\ \ref{fig:ch1_cosmic_timeline} shows a
sketch of the main cosmic epochs that are treated in this textbook.
Soon after the Big Bang, the baryonic matter was fully
ionised and coupled with a `bath' of black-body photons. In this
photon--baryon fluid, the Universe was opaque because photons
could not propagate freely due to the incessant Thomson scattering with free
electrons. As the Universe expanded, its temperature and density gradually decreased
and, about three minutes after the Big Bang,
the nuclei of elements heavier than hydrogen (basically only helium and lithium)
formed through a process called primordial nucleosynthesis. About 400\,000 years
after the Big Bang, the temperature and density dropped enough to allow lithium,
helium and hydrogen to gradually recombine with electrons and form neutral atoms.
This phase is called cosmological recombination. This is the epoch when the Universe
became transparent because photons started to propagate freely thanks to the negligible
role of Thomson scattering. The CMB radiation observed in the present-day Universe
was originated in this phase and therefore represents the earliest possible image
of the Universe. After recombination, the Universe was
filled of dark matter and diffuse neutral gas composed of hydrogen, helium and
lithium only. It is from the evolution of this primordial gas that the first luminous
objects and galaxies began to form.

Understanding galaxy formation and evolution is a complex task because it involves
several physical processes, their mutual interactions, and their evolution as a
function of cosmic time. This is one of the most multi-disciplinary areas of
astrophysics as it requires the cross-talk among a wide range of fields such as
cosmology, particle physics (including dark matter) and the physics of baryonic
matter. Galaxy formation and evolution is also a relatively
young research field because galaxies were recognised as such only about a century ago,
and their observation at cosmological distances became possibile only in the mid-1990s
thanks to the advent of ground-based 8--10 m diameter telescopes in the optical and
near-infrared spectral ranges, in synergy with the {\em Hubble Space Telescope} ({\em HST}).

\looseness=-1 The first step in the study of galaxy formation and evolution requires the definition
of a cosmological framework (currently the $\Lambda$CDM model) within which galaxies
form and evolve. The second step is to include the formation and evolution
of dark matter halos which will host the first luminous objects and galaxies.
In the $\Lambda$CDM model, dark matter halos are the results of the gravitational
collapse of CDM in the locations where the matter~density is high enough to
locally prevail over the expansion of the~Universe. As a~matter of fact, the
competition between the expansion of the Universe and gravity is one~of the key
processes in galaxy formation.
On the one hand, if we take a large volume of the Universe at a given time, the mean
matter density decreases with increasing cosmic time~due to the expansion
of the volume itself. On the other hand, the masses present in the same volume
attract each other due to the reciprocal gravitational forces. In the early Universe,
the typical masses of these halos were small, but they subsequently
grew hierarchically with cosmic time through the merging with other halos and
with the accretion of diffuse dark matter. Part of the gas is expected to follow the
gravitational collapse of dark matter halos, and then to settle into their potential wells.
The possibility to form a galaxy depends on whether this gas can have a rapid gravitational
collapse. First of all, gravity must prevail over the internal pressure of the gaseous matter.
However, this is not
sufficient because the temperature rises as soon as the contraction proceeds. Gas
heating is the enemy of galaxy formation because it increases the
internal pressure and hampers gravitational collapse. This is why
the second key requirement for galaxy formation is that gas cooling
prevails over heating.
Gas cooling can be produced by the emission of continuum radiation and
spectral lines. The emitted photons abandon the gas cloud, carrying energy away, and
therefore making the gas cooler and more prone to further gravitational collapse.


The cosmic epoch before the formation of the first collapsed objects (known as
first stars or Population III stars) is named the dark ages
because the Universe was made only of neutral gas, and luminous sources were completely
absent ({Fig.}\ \ref{fig:ch1_cosmic_timeline}). We think that Population III stars began to form about 100 million years
after the Big Bang from the collapse of pristine gas (H, He, Li) within dark matter
halos with masses around $10^{6}$ $\mathcal{M}_{\odot}$. At these early epochs, the main
radiative coolants of the gas were primordial molecules such as LiH, HD and H$_2$
previously formed through gas-phase chemical reactions. This collapse led to the formation
of protostellar objects and the subsequent ignition of the first thermonuclear reactions in
the cores of Population III stars. When these systems started to shine, their strong
ultraviolet radiation photoionised the surrounding gas. This was the beginning of the
reionisation era. Population III stars ended their life very rapidly and vanished
with the expulsion of most of their gas from their dark matter halos by violent supernova
explosions. Thus, having lost most of the initial gas, these halos could not host
further episodes of star formation. It is therefore thought that the formation of
the first galaxies occurred later (a few hundred million years after the
Big Bang) in larger dark matter halos with masses around $10^{8}$ $\mathcal{M}_{\odot}$.
These objects are called galaxies because they
were massive enough to gravitationally retain a substantial fraction of the
gas to prolong star formation without losing and/or heating it excessively
due to supernova explosions.

\begin{figure}[!b]
{\includegraphics{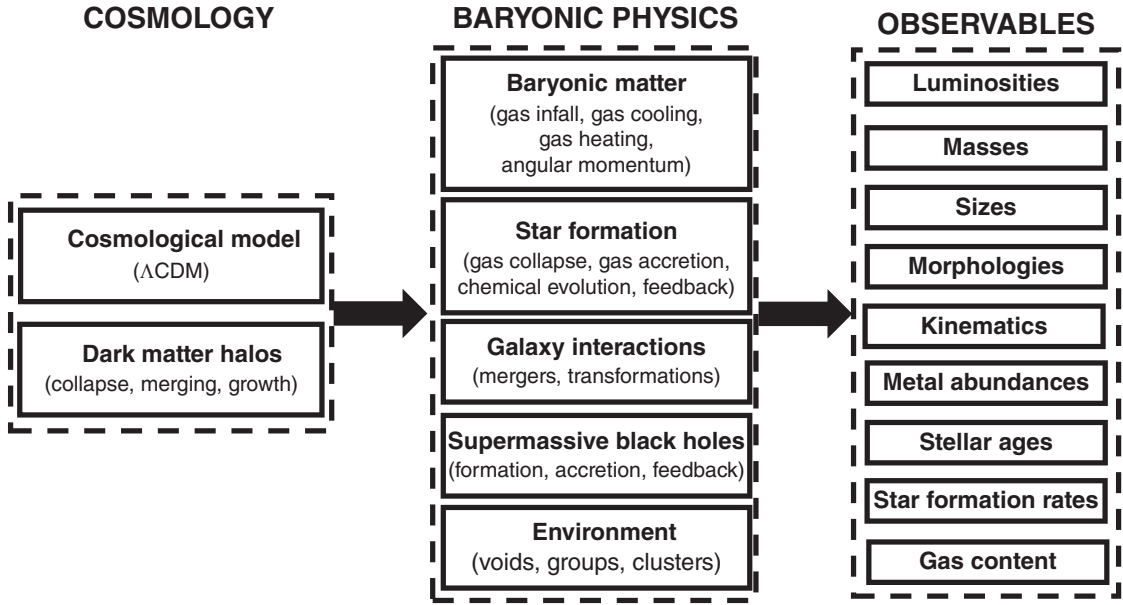}}
{\caption{The main ingredients of models of galaxy formation and evolution.
{\em Left}. The cosmological
model and the properties of dark matter halos define the `skeleton' within which
galaxies form and evolve. {\em Centre}. The main processes that drive the evolution
of baryonic matter and galaxy formation. {\em Right}. The predicted properties
of galaxies that are compared with the observations to verify the reliability
of theoretical models.}\label{fig:intro_flowchart}}
\end{figure}

After these early phases, galaxy formation proceeded
following a wide range of evolutionary paths depending on the local conditions,
the properties of the gas and the interactions with other systems (e.g.\ merging
of their host halos). This is why the full understanding of galaxy formation and evolution
is complex and requires a self-consistent treatment of the physical processes
of baryonic matter (gas, stars and dust), their kinematics, their evolution within an
expanding Universe, and the gravitational interactions with
the dark matter component. The physics of baryonic matter is particularly complicated
as it involves a variety of ingredients such as radiative processes, multi-phase gas
physics and dynamics, gas cooling and heating, radiative transfer,
star formation, stellar evolution, metal enrichment and feedback.
Moreover, galaxy evolution involves the formation of supermassive black holes,
the associated accretion of matter, and also the consequent feedback processes on the surrounding
environment. A further complication is that all these processes and their evolution
must be investigated on very wide ranges of spatial scales (from sub-parsec to megaparsec)
and timescales, say from the lifetime of the most massive stars ($\sim 10^6$ yr) to
the age of the Universe ($\sim 10^{10}$ yr).
{Fig.}\ \ref{fig:intro_flowchart} illustrates the main ingredients that need to be
included for the physical description of galaxy formation and evolution.

\section{Galaxies: Near and Far, Now and Then}
\label{sec:chap1_galaxies_near_far}

Given the above complexity, how can we study galaxy formation and evolution?
One approach is through theoretical models which describe coherently the physical
processes involved in the formation of galaxies and their subsequent evolution from
the smallest to the largest scales. In these models, the $\Lambda$CDM cosmology
framework provides the initial conditions (e.g.\ the dark and baryonic matter
fractions, the expansion rate of the Universe as
a function of time, the properties of CDM halos and the hierarchical
evolution of their masses). Once the cosmological framework is defined, galaxies can
be modelled with two main methodologies. The first is based on cosmological hydrodynamic
simulations, which follow as much as possible self-consistently the evolution of
gas, star formation and feedback processes within dark matter halos.
These simulations are very time consuming. This implies that sub-galactic
scales can be simulated at the price of not covering large volumes of the Universe
due to the limited computational resources. The second approach, called
semi-analytic, consists in treating the physics of baryonic matter with a
set of analytic prescriptions that, combined with the theoretically predicted evolution
of dark matter halos, are tuned to reproduce the observed properties of present-day galaxies.
The semi-analytic approach is cheaper from the computational
point of view and therefore allows one to simulate large volumes of the Universe
up to gigaparsec scales. However, the price to pay is that only the global properties of
galaxies can be studied, and limited spatially resolved information is available.
For these reasons, the two methods are complementary to each other.
A further possibility is to perform analytic/numerical
modelling of specific processes which take place within galaxies. An example is
given by the chemical evolution models applied to the Milky Way.

The other approach to study galaxy formation and evolution is complementary
to the theoretical modelling, and consists in the direct observation of galaxies in
order to obtain data (images and spectra) from which the physical and structural
properties can be extracted.
A first possibility is the so-called archaeological approach where
present-day galaxies are exploited as `fossils'\footnote{As present-day galaxies are considered `fossils',
the archaeological approach should be more appropriately called `palaeontological'.}
from which it is possible to reconstruct their past history based on what is observed today.
For instance, the ages and metal abundances of the stellar populations present
in a galaxy allow us to infer how star formation and the enrichment of heavy
elements evolved as a function of
cosmic time. With this approach, the most reliable results are obtained
when the stars within a galaxy can be observed individually and therefore
can be placed on the Hertzsprung--Russell diagram. Unfortunately,
with the current telescopes, this can be done only within the Milky Way and
for galaxies in the Local Group, a $\approx$1 Mpc size region where the Galaxy
is located together with its neighbours.
The study of our Galaxy is so important as a benchmark of galaxy evolution
studies that the $Gaia$ space mission has been designed to obtain distances and
proper motions of more than a billion stars, with radial velocity measurements
for a fraction of them. $Gaia$ allowed
us to derive a kinematic map of our Galaxy that is
essential to investigate its formation and evolution. Beyond the Local Group,
galaxies become rapidly too faint and their angular sizes are too small to observe
their stars individually. In these cases, one has to rely on the `average'
information that can be extracted from the so-called integrated light,
i.e.\ the sum of the radiation emitted by the entire galaxy (or by a region of it).

Besides the archaeological studies in the present-day Universe, galaxy formation
and evolution can also be investigated with the so-called look-back approach.
This consists in the observation of galaxies at cosmological distances.
Since light travels at a finite speed, the photons emitted from more distant
galaxies reach us after a longer time interval.
This means that distant galaxies appear today to us as they were in the past.
Thus, it is possible to observe directly the evolution of galaxy properties if
we observe galaxies at increasing distances. The fundamental assumption that makes
the look-back time approach possible is that the Universe is homogeneous on large scales,
so the global properties of the galaxy population on sufficiently large volumes
are independent of the position in the Universe. This implies that
galaxies in the local volume, in which our Galaxy is located, are representative
of the general population of present-day galaxies. Similarly the galaxies observed
in a distant volume are assumed to be representative of the past population of galaxies.
For instance, if we want to investigate the evolution of spiral galaxies,
we need to observe samples of this type of galaxies at increasing
distances (i.e.\ larger look-back times) and to study how their properties (e.g.\
size, rotation velocity, mass, star formation) change with cosmic
time. With this approach, it is truly possible to trace the detailed evolution
of galaxies billions of years ago.


The archaeological and look-back approaches are complementary to each
other, and their results are essential to build theoretical models and
verify their predictions. However, in both cases multi-wavelength data
are needed to provide a
complete view of galaxy properties and their evolution.
The reason is that galaxies are multi-component systems which emit radiation
in different regions of the electromagnetic spectrum through diverse processes.
For instance, due to
the typical temperatures of the stellar photospheres, the
starlight is concentrated from the ultraviolet to the near-infrared. Instead,
the study of the interstellar molecular gas and dust requires observations
from the far-infrared to the millimetre, the atomic hydrogen must be investigated
in the radio, and the hot gas and the supermassive black hole activity in the
ultraviolet and X-rays. The
multi-wavelength approach is limited by the terrestrial atmosphere
which is opaque and/or too bright in several spectral ranges. Ground-based telescopes can
observe only in the optical, near-infrared and in a few transparent windows
of the submillimetre, millimetre and radio. The other spectral ranges are accessible with
space-based telescopes. The major advance in multi-wavelength
studies of galaxy evolution at cosmological distances became possible thanks
to the concurrence of
ground- and space-based telescopes which, for the first time, allowed the
identification of galaxies at cosmological distances. In the realm of space telescopes,
the main contributions to galaxy evolution studies have come from the {\em Chandra
X-ray Observatory}, {\em XMM-Newton} (X-rays), {\em Galaxy Evolution Explorer} (\textit{GALEX};
ultraviolet), {\em HST} (optical/near-infrared), {\em Spitzer} (mid-infrared) and {\em Herschel} (far-infrared). In ground-based observations,
the look-back approach became a reality with the advent of the \hbox{$8$--$10$ m} diameter
{Keck} telescopes and the {Very Large Telescope} (VLT) operating since the \hbox{mid-1990s}
in the optical and near-infrared, followed by other facilities of comparable
size ({Gemini}, {Subaru}, {Gran Telescopio Canarias} and the {South
African Large Telescope}). The {James Clerk Maxwell Telescope}
(JCMT), the telescopes of the {Institut de Radioastronomie Millim\'etrique} (IRAM)
(such as the {NOrthern Extended Millimeter Array}; NOEMA),
the {Atacama Large Millimetre Array} (ALMA) and the {Karl G. Jansky
Very Large Array} (VLA) were essential in opening new windows on galaxy
evolution at submillimetre, millimetre and radio wavelengths.

The multi-wavelength data provided by these facilities allow us to study galaxy
evolution. Imaging observations are crucial to study the morphology and
structure of galaxies and their relations with their physical properties.
Spectroscopy provides information on the stellar populations, interstellar matter and
the presence of supermassive black holes through the analysis of continuum spectra
and spectral lines. Furthermore, the Doppler effect allows us to derive the galaxy
kinematic properties, to measure dynamical masses and to study scaling relations.
Last but not least, the collection of large galaxy samples over wide sky areas allows
us to use galaxies as luminous markers to trace the spatial distribution of the
underlying dark matter and to exploit them as cosmological probes.

\section{Galaxies: the Emerging Picture and the Road Ahead}
\label{sec:chap1_emerging_picture}

What is the picture emerging from the synergy of observations and theory?
Most studies seem to converge on a scenario in which the evolution of
galaxies is driven across cosmic times by the so-called baryon cycle. Galaxies
are thought to accrete gas from the surrounding environment and gradually convert
it into stars. The cooling and condensation of neutral hydrogen, and its conversion
into molecular hydrogen to fuel star formation, are therefore key processes driving
galaxy evolution. In this picture, galaxies grow mainly through gas accretion from the
intergalactic medium, while there is a complex equilibrium between gas inflow,
the conversion of
the available gas reservoir into stars, and gas ejection and heating by feedback processes.
In parallel, supermassive black holes form at the centres of galaxies, and trigger
the temporary phase of an active galactic nucleus (AGN) whenever the accretion of cold
material is efficient enough. Part of the stellar mass is lost by stars during their
evolution through winds, planetary nebulae, novae and supernovae. This ejected mass
seeds the interstellar medium with metals, molecules and dust grains, while starburst
winds and jets from AGNs provide feedback and launch gas outflows. Metal-enriched and
pristine halo gas eventually cools and accretes onto the disc to form new stars and feed
the central black hole, starting the cycle again. A complex interplay is therefore
expected among these processes as a function of galaxy properties, environment and
cosmic time. Hence, understanding the evolution of the baryon cycle has become a key
question that must be addressed to shed light on the critical steps of galaxy formation
and evolution.

Despite the major progress in this research field, the overall picture
is still largely incomplete, and several key questions are still open. However,
new facilities operating in space such as the
\textit{James Webb Space Telescope} (\textit{JWST}), {\em Euclid}, the {\em Wide Field Infrared Survey
Telescope} (\textit{WFIRST}) in the optical/infrared, and the {\em Athena X-ray Observatory} in the X-rays have been designed to open
new windows through the identification and multi-wavelength studies of galaxies
across the entire range of cosmic times since the end of the dark ages.
In this landscape, a key role is played by the synergistic studies done with the new
generation of gigantic telescopes on the ground such as the {Extremely Large
Telescope} (ELT), the {Giant Magellan Telescope} (GMT) and the {Thirty Meter Telescope} (TMT) in the optical/near-infrared,
and the {Square Kilometre Array} (SKA) in the radio.
In addition, the development of new numerical models
and improved supercomputing facilities allow us to perform new simulations that are
essential to investigate how the complex physics of baryonic matter and its interplay
with the dark matter halos drove the evolution of different galaxy types as a function of
cosmic time.

%





\end{document}